\documentclass[review,12pt]{elsarticle}

\usepackage{float}
\usepackage{amssymb}
\usepackage{amsmath}
\usepackage[T1]{fontenc}
\usepackage[colorlinks]{hyperref}
\usepackage{amsxtra}
\usepackage{amstext}
\usepackage{amssymb}
\usepackage{graphicx}
\usepackage{amsfonts}
\usepackage{multicol}
\usepackage{setspace}
\newtheorem{thm}{Theorem}[section]

\newtheorem{defn}[thm]{Definition}

\journal{Journal of Theoretical Biology}

\begin{document}

\begin{frontmatter}

\title{Forest structure in epigenetic landscapes}

\author{Yuriria Cortes-Poza}
\ead{yuriria.cortes@cimat.mx}
\address{CIMAT-M\'{e}rida \\ Parque Cient\'{\i}fico y Tecnol\'{o}gico de Yucat\'{a}n, Sierra Papakal, Yucat\'{a}n}

\author{J. Rogelio Perez-Buendia}
\ead{rogelio.perez@cimat.mx}
\address{CONACyT-CIMAT, M\'{e}rida \\ Parque Cient\'{\i}fico y Tecnol\'{o}gico de Yucat\'{a}n, Sierra Papakal, Yucat\'{a}n}

\maketitle

\begin{abstract}

Morphogenesis is the biological process that causes the emergence and changes of patterns (tissues and organs) in living organisms. It is a robust, self-organising mechanism, governed by Genetic Regulatory Networks (GRN), that hasn't been thoroughly understood. In this work we propose \emph{Epigenetic Forests} as a tool to study morphogenesis and to extract valuable information from GRN. Our method unfolds the richness and structure within the GRN.

As a case study, we analyze the GRN during cell fate determination during the early stages of development of the flower \textit{Arabidopsis thaliana} and its spatial dynamics. By using a genetic algorithm we optimize cell differentiation in our model and correctly recover the architecture of the flower.

\end{abstract}


\end{frontmatter}

\newpage
\section{Introduction}
\label{sec:intro}

Genetic Regulatory Networks (GRN) govern morphogenesis, the process responsible for producing the complex shape of full grown organisms from a few cells (\cite{Keller-Morph}, \cite{Ettensohn-Morph}) and every process of life including metabolism, the cell cycle and signal transduction.
Genetic Regulatory Networks (GRN) are a composed of a collection of genes that interact between each other and with external factors, forming a complex network (\cite{Davidson-GRN}). 

In recent years GRN have gained a lot of attention and great advances have been made (\cite{Karlebach-GRN}, \cite{Emmert-GRN}). From cancer identification (\cite{Iyer-Cancer}),(\cite{Chang-Cancer}) and diabetes (\cite{Jesmin-Diabetes}) to the formation of animals' bodies (\cite{Davidson-Animal}) GRN are being used to model and understand diverse biological processes. However these complex networks haven't been thoroughly understood and there is a lot of experimental data that needs to be analyzed using computational and mathematical techniques.

In this work, we propose \textit{Genetic Regulatory Trees}, as a way to study Genetic Regulatory Networks. By modeling GRN as a discrete dynamical system (boolean automata) and analyzing all its possible states, we find that they have a forest structure, where each tree corresponds to a fixed state of the automata. This forest models Waddington's epigenetic landscape of the network (a developmental model that illustrates the mechanics of cell fate differentiation see section (\ref{sec:2.1})). The use of this technique to model GRN and its epigenetic landscape is a novel approach.

As a case study we analyze the GRN of the flower \textit{Arabidopsis thaliana} during cell fate determination. In (\cite{Alvarez-FM}), using experimental data, we obtained the gene regulatory network (GRN) of the flower \textit{Arabidopsis thaliana} that determines the fate of floral organ cells.
The network models the interactions between genes responsible for cell differentiation: each node in the automata represents a certain gene, the edges are the interactions between them and the logical rules that govern the automata are based on detailed experimental data.

Genetic Regulatory Trees are defined and built as a directed graphs that turn out to have a directed rooted tree structure. Each node in a tree represents a given state of the Boolean automata (GRN) and every possible state is considered. The edges link states that follow each other in the dynamical system and the root of each tree represents a steady state of the automata, so that there will be as many trees as steady states. Once the trees are built, we define chains of cell types that will traverse the trees in a certain order. Each link in the chain will represent a group of undifferentiated cells (of the meristem) with the same genetic configuration, and neighbouring links will be genetically similar to each other. Biologically we can think of it as a radial longitudinal section of the meristem. We define an energy measurement of a chain (the energy it requires for all of its links to differentiate). Considering nature is efficient, we use Genetic Algorithms to minimize this energy and expect to recover accurately the spatial configuration of the flower during cell fate determination. The set of trees represent the epigenetic landscape that models how the different environmental and genetic forces affect cellular differentiation.

This method is general enough to be used in any GRN as a tool to retrieve very interesting information about them, like the robustness of cell differentiation into each flower organ (trees), the importance of a specific state (a node in the tree), and the underlying dynamics of certain processes, like cellular differentiation, as studied in this work.

This paper is organized in the following manner: The first section is this introduction; on section two we present biological background information and we detail the discrete dynamical system (the Boolean automata) of the GRN we will analyze. On the third section we present the details of our model and we construct our Genetic Regulatory Trees taking into account experimental data (from the discrete dynamical system). On section 4 we present the details of the genetic algorithm, used to minimize the energy of the paths, and the obtained results. Finally on the last section we discuss our concluding remarks.

\section{Background information}

It this section we provide the biological information needed to understand our work and describe the Genetic Regulatory Network model we are using.

\label{sec:2}
\subsection{Biological background}
\begin{figure}[H]
\centering
  \includegraphics[width=8cm]{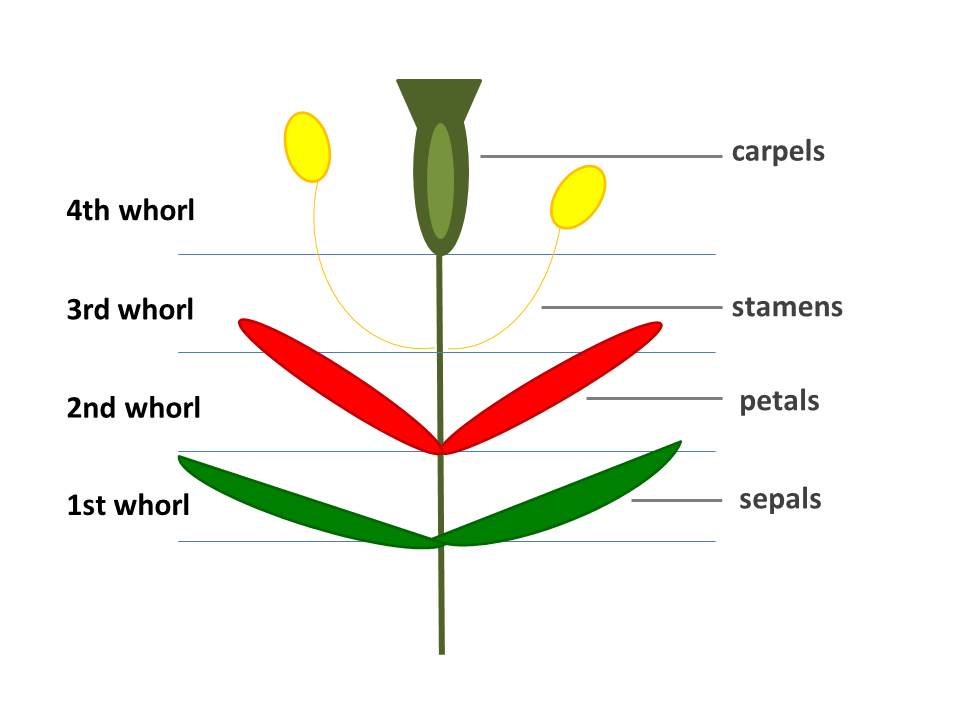}\\
  \caption{Whorls (circular sections of the flower each consisting of a different organ) of a typical Angiosperm: sepals in the first whorl, petals in the second one, stamens in the third one and carpels in the center.}\label{fig:whorls}
\end{figure}

The flower organs of all species of Angiosperms (approx. 250,000) are organized in four concentric rings (whorls), which are, from the outer rim to the center: sepals, petals, stamens and carpels (fig (\ref{fig:whorls})). The only known exception to this configuration is the one observed in the flower \textit{Lacandonia schismatica} where the position of its stamens and carpels is inverted.

We work with the flower of the plant \textit{Arabidopsis thaliana}. This plant was the first one whose complete genome was sequenced and has been extensively studied (\cite{Arabidopsis-Gen}), (\cite{Alvarez-Arabidopsis}).

In (\cite{Alvarez-FM}), the gene regulatory network (GRN) of the flower \textit{Arabidopsis thaliana} was obtained. It is based on experimental data and it's defined as a Boolean automata which determines the fate of floral organ cells.

Using this model, we define and construct our Genetic Regulatory Trees as we will see in the following section.

\subsection{Epigenetic landscape}
\label{sec:2.1}
Epigenetic landscapes, originally proposed by Waddington in 1975 (\cite{Waddington-EL}), are developmental models that illustrate the mechanics of cell fate differentiation. These models use as a metaphor a mass in a potential field with a certain number of basins of attraction and paths that lead to each one of them. We propose a way to model the epigenetic landscape associated with a certain GRN using Genetic Regulatory Trees.
As a case study we use the GRN of the flower \textit{Arabidopsis thaliana} during cell fate determination. Each tree will correspond to a different flower organ (sepals, petals, stamens, carpels and the inflorescences). We use the information obtained from the Boolean network, a discrete dynamical system  to build them. In the following section we present the details of this network.

\subsection{Boolean network}
\label{sec:2.2}
In (\cite{Alvarez-GRN}) a discrete dynamical system was used to explore the dynamics of cell fate determination during the early stages of flower development. The system is a Boolean automata, consisting of $13$ nodes, each one corresponding to a specific gene \footnote {The genes considered are: FUL (Fruitfull), FT (Terminal flower), AP1 (Apetala 1), EMF1 (Embrionic flower 1), LFY (Leafy), AP2 (Apetala 2), WUS (Wuschel), AG (Agamous), TFL1 (Terminal flower 1), P1 (Piscilata 1), SEP (Sepallata), AP3 (Apetala 3) and UFO (Unusual flower organ) \cite{Alvarez-Arabidopsis}.}.

\begin{figure}[H]
\centering
  \includegraphics[width=8cm]{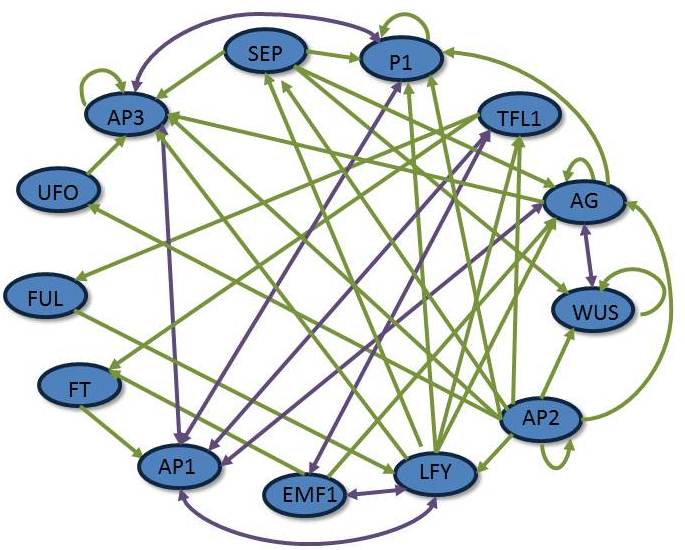}\\
  \caption{Directed graph that represents the celular automata: each edge is a gene and the edges represent interactions between them.}\label{Fig:grafAut}
\end{figure}

Each node has two possible states ($0$ or $1$) that are updated according to experimentally obtained rules, that correspond to the interaction between genes. These interactions can be expressed as a directed graph as shown in figure (\ref{Fig:grafAut}), where we can see a green edge from node $N_i$ to node $N_j$ if node $Ni$'s state affects $N_j$ and a purple edge between node $N_i$ and $N_j$ if both nodes affect each other. The rules that the automata follows (obtained experimentally) can be found in (\cite{Alvarez-GRN}).

There are $2^{13}$ possible initial conditions (binary numbers form $0$ to $2^{13}-1$). The system is iterated, starting from each initial condition. It converges to eight different attractors \footnote{For simplicity we don't take into account two additional attractors that are irrelevant in this work because they correspond to flower organs that are already considered (petals and stamens) and the number of initial conditions that reach them is very small.}. Every single initial condition lands in one of these ten states. Each atractor represents one of the main cell types observed during the early stages of flower development (the meristematic cells of the inflorescence and the primordial cells of the flower meristems of sepals, petals, stamens and carpels) \cite{Espinosa-GRN}. Each equilibrium point has thirteen components (see table(\ref{tab:1})).

\begin{table}
\caption{Equilibrium points: each component of the string $q_i$ represents the state of a given gene (node): 0 is it is inactive and 1 if it is active. There are 13 characters in each $q_i$, one for each node.}
\centering
\label{tab:1}
\begin{tabular}{|l|c|}
\hline
 \multicolumn{2}{|l|}{Floral organ\qquad\qquad\qquad\ \ \ Atractor} \\
 \hline
   Inflorescence 1         & $p_{I_1}=[0,0,0,1,0,0,0,0,1,0,0,0,0]$\\
   Inflorescence 2         & $p_{I_2}=[0,0,0,1,0,0,0,0,1,0,0,0,1]$\\
   Inflorescence 3         & $p_{I_3}=[0,0,0,1,0,0,1,0,1,0,0,0,0]$\\
   Inflorescence 4         & $p_{I_4}=[0,0,0,1,0,0,1,0,1,0,0,0,1]$\\
   Sepals                  & $p_S=[0,1,1,0,1,1,0,0,0,0,1,0,0]$\\
   Petals                  & $p_P=[0,1,1,0,1,1,0,0,0,1,1,1,1]$\\
   Stamens                 & $p_T=[1,1,0,0,1,1,0,1,0,1,1,1,1]$\\
   Carpels                 & $p_C=[1,1,0,0,1,1,0,1,0,1,1,0,0]$\\
   \hline
    \end{tabular}
\end{table}
\ \newline

The number of nodes (initial conditions land in each fixed point) in each tree will be the following:
\begin{eqnarray}\label{eq:bn1}
   c_S=152, c_P=160, c_T=3744, c_C=3608, \\
   c_{I_1}=128, c_{I_2}=128, c_{I_3}=64, c_{I_4}=64.
\end{eqnarray}

where the subscripts S, P, E and C correspond to sepals, petals, stamens and carpels respectively and $I_i$ to the inflorescences.  (\cite{Espinosa-GRN},\cite{Alvarez-FM}).

Based on this discrete model, we construct the Genetic Regulatory Trees.

\section{Our method}
\label{sec:3}

\subsection{Genetic Regulatory Trees}
Genetic Regulatory Trees model the epigenetic landscape of Genetic Regulatory Networks and are constructed by analyzing the states of the discrete model detailed in the previous section.

\begin{figure}[H]
\centering
  \includegraphics[width=14cm]{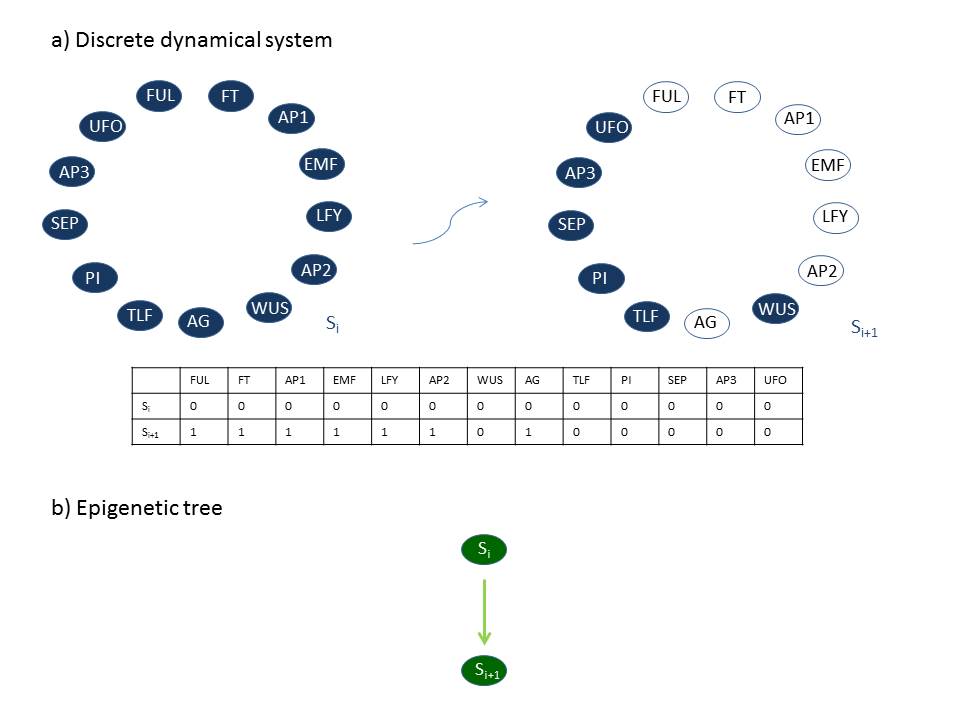}\\
  \caption{Construction of the tree from the dynamical system. (a) Two different states of the dynamical system, $S_i$ leads to $S_{i+1}$ (b) Two nodes in a tree, first node corresponds to $S_i$ and the second one to $S_{i+1}$.}\label{Fig:construcTree}
\end{figure}

The Genetic Regulatory Trees are directed graphs that turn out to have a directed rooted tree structure. To build them we proceed as follows:
Each node $S_i$ in the tree will correspond to a specific state of the discrete dynamical system detailed above. Every possible state will be considered (in our case there will be a total of $2^{13}$ nodes in the trees). There will be a directed edge from node $S_i$ to $S_j$ of the tree if (and only if), following the rules of the automata, the state of node $S_i$ leads to the one corresponding to node $S_j$ (see fig. (\ref{Fig:construcTree})).

Proceeding this way until every state is analyzed, we obtain an in-tree (orientated towards the root) for every fixed state in the discrete dynamical system. The root of each tree will be precisely a fix state, that in our case represents a flower organ. The initial points on the dynamical system, that is, the points which are not the iterate of any other point will be the leafs of the trees.

Since our dynamical system has $8$ fixed states, we will end up with $8$ different trees, which can be seen in the following graphs:

\begin{figure}[H]
\centering
  \includegraphics[width=8cm]{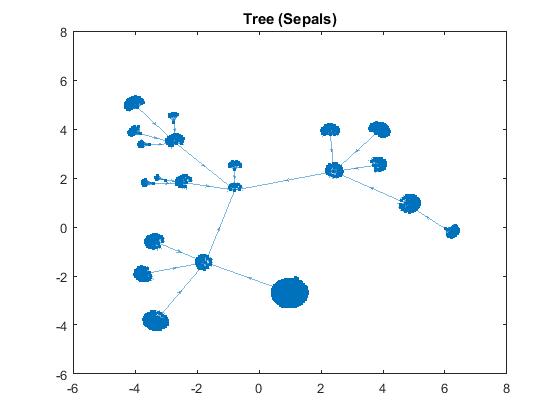}\\
  \caption{The tree corresponding to Sepals}\label{Fig:TreeS}
\end{figure}

\begin{figure}[H]
\centering
  \includegraphics[width=8cm]{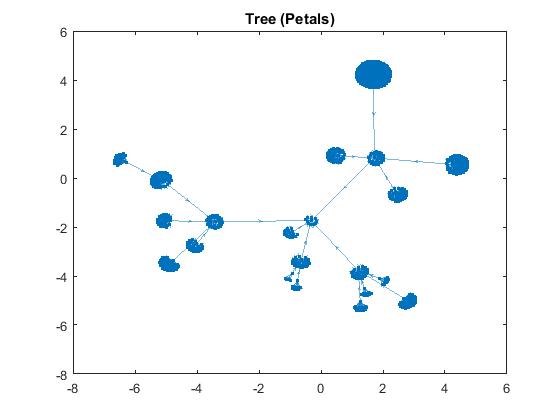}\\
  \caption{The tree corresponding to Petals}\label{Fig:TreeP}
\end{figure}

\begin{figure}[H]
\centering
  \includegraphics[width=8cm]{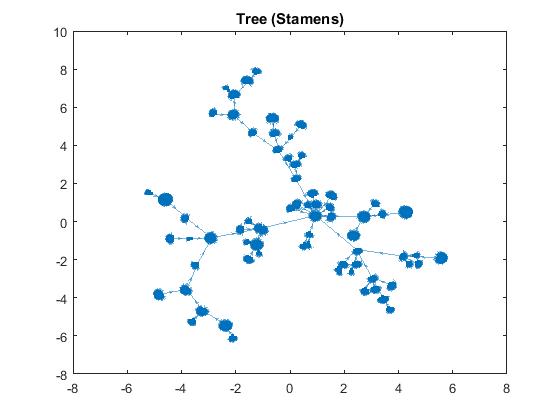}\\
  \caption{The tree corresponding to Stamens}\label{Fig:TreeT}
\end{figure}

\begin{figure}[H]
\centering
  \includegraphics[width=8cm]{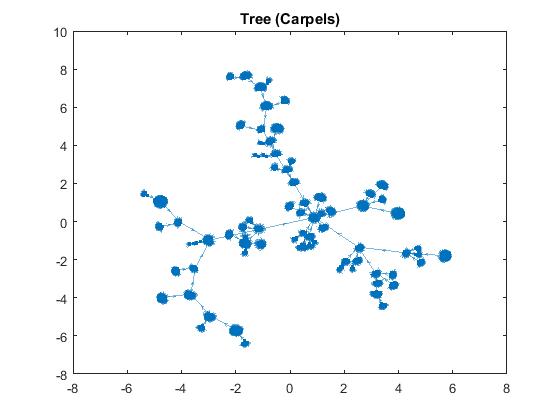}\\
  \caption{The tree corresponding to Carpels}\label{Fig:TreeC}
\end{figure}

\begin{figure}[H]
\centering
  \includegraphics[width=8cm]{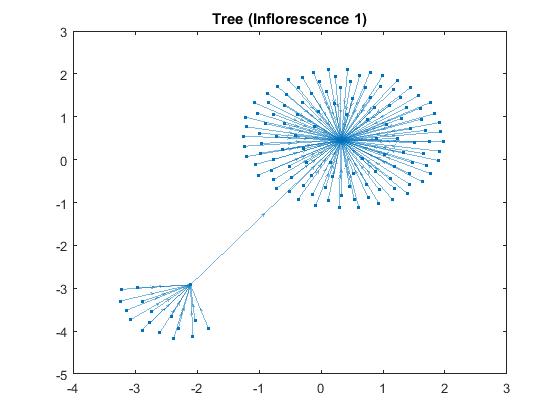}\\
  \caption{The tree corresponding to Inflorescence 1}\label{Fig:TreeI1}
\end{figure}

\begin{figure}[H]
\centering
  \includegraphics[width=8cm]{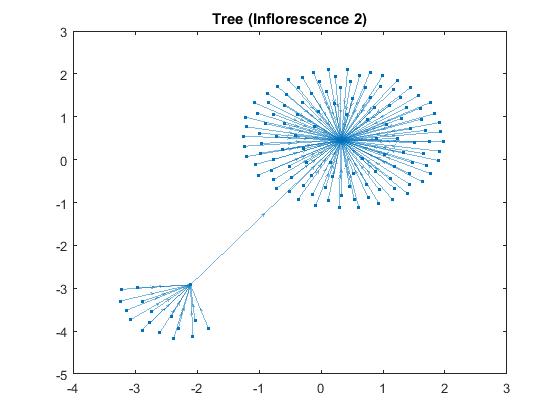}\\
  \caption{The tree corresponding to Inflorescence 2}\label{Fig:TreeI2}
\end{figure}

\begin{figure}[H]
\centering
  \includegraphics[width=8cm]{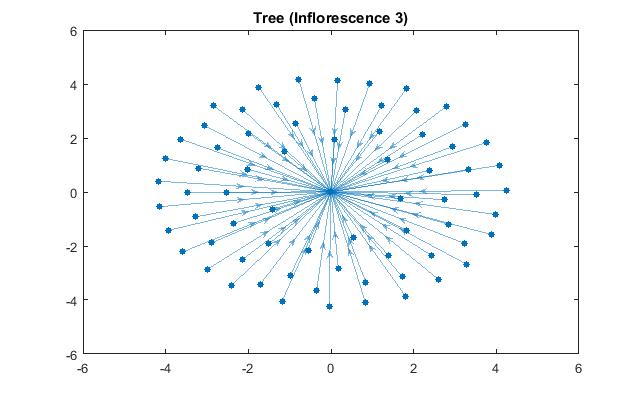}\\
  \caption{The tree corresponding to Inflorescence 3}\label{Fig:TreeI3}
\end{figure}

\begin{figure}[H]
\centering
  \includegraphics[width=8cm]{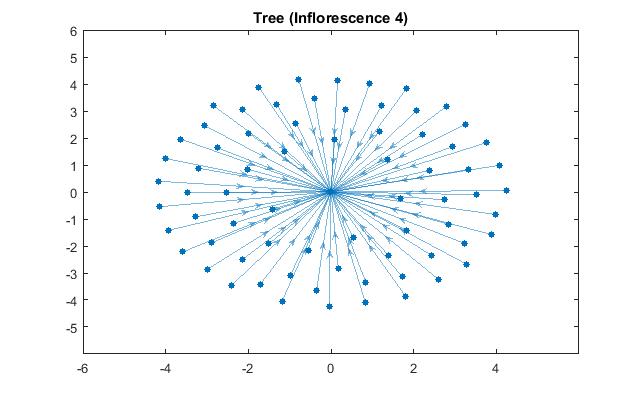}\\
  \caption{The tree corresponding to Inflorescence 4}\label{Fig:TreeI4}
\end{figure}

These trees represent the epigenetic landscape of the Genetic Regulatory Network. As we can see, their structure is very interesting.

\subsection{Chains of cell types}

We are now ready to define chains $C$ of cell types. A chain will be formed by $L$ links, each of which will represent a group of undifferentiated cells with the same genetic configuration (a possible state of the automata), that is, an array with $13$ entries (one for each gene) that can be in one of the two possible states we are considering ($0$ or $1$). We'll construct our chains so that the first link will be a node on the tree corresponding to sepals and neighbouring links will be genetically similar to each other.

\begin{figure}[H]
\centering
  \includegraphics[width=8cm]{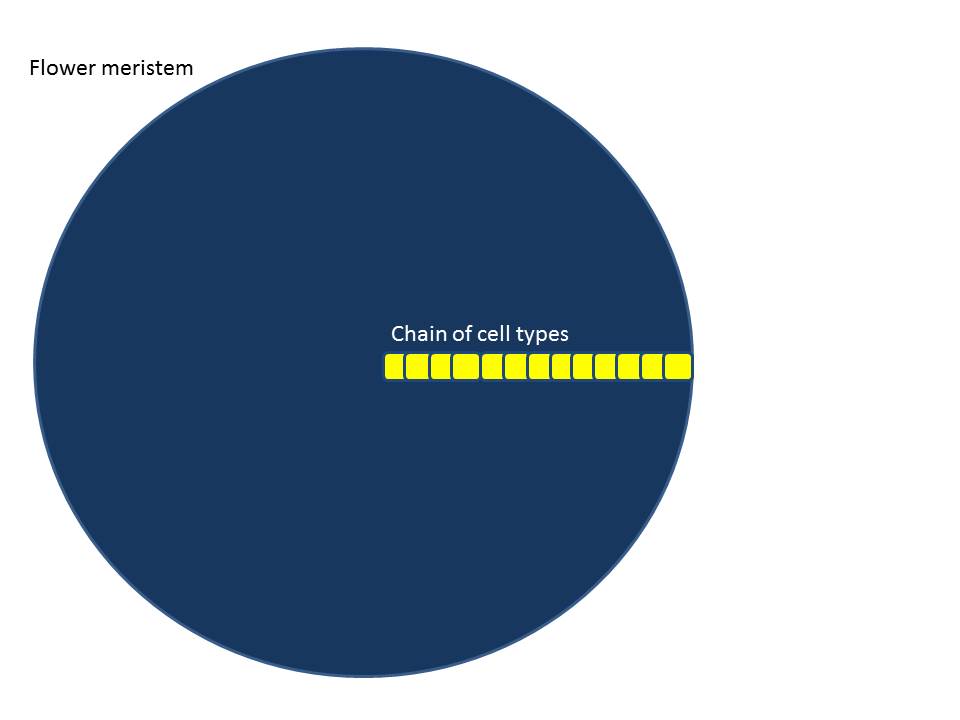}\\
  \caption{Chain of cell types taken from the meristem of the flower}\label{Fig:chain}
\end{figure}

Biologically, we can think of a chain as a radial longitudinal section of the meristem (recall that flowers are radially symmetrical) (see figure(\ref{Fig:chain})).

The formal definition is the following:

\begin{defn}(Undifferentiated cells chain)
A chain of undifferentiated cell types will consist of an array of $L$ links,
$$C=[c_1,c_2,\ldots c_L],$$
where each link is a possible state of the automata, that is, $c_i=[x_1,x_2,\ldots,x_{13}]$ and $x_i=\{0,1\}$. The initial one, $c_1$ will be a node in the starting tree (sepals) and each consecutive link will be formed by modifying randomly one (and only one) of the entries of the previous one.
\end{defn}
This construction guarantees that neighboring cell types in the chain will be \textit{similar} to each other. To measure how similar two cells are, we will use Hamming distance:

\begin{defn}[Hamming-distance]
Given two strings of the same length $s_i, s_j$, the Hamming distance between them $d(s_i,s_j)$ is the number of positions at which the corresponding symbols differ (it measures the minimum number of substitutions needed to change one string into the other).
\end{defn}

As we can see, Hamming distance between any two neighboring cells in a chain will be one. Note that a cell type (link) in a chain belongs to a given tree (whose root is a steady state of the automata) if it is an initial condition of its steady state. Since in our case, there are nodes in different trees whose distance between them is one, a chain can have links in different trees. We say that a chain traverses a certain number of trees.

Also note that trees are acyclic graphs, so that there will be at most one directed path from any one node to another one, in particular there will (always) be a unique path between from node $c_i$ to the root of the tree. 

\begin{defn}[Specialization path]
The path from a given node $c_i$ in a tree to its root is called the specialization path of $c_i$ and we denote it as $SP(c_i)$.
\end{defn}

A chain specializes (differentiates) $C=[c_1,c_2,\ldots, c_L]$ when each of its links reaches a steady state (following the rules of the automata), that is if $c_i$ belongs to tree $T_j$, it must traverse its specialization path $SP(c_i)$ reaching this way, the root of $T_j$.

\begin{figure}[H]
\centering
  \includegraphics[width=8cm]{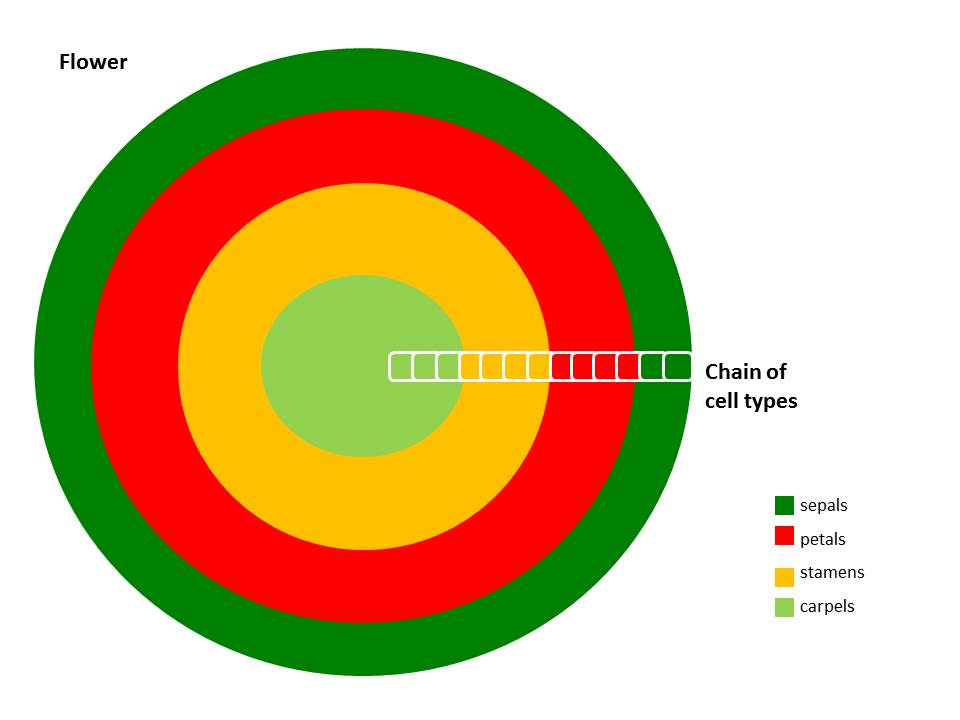}\\
  \caption{Chain of differentiated cell types in the flower}\label{Fig:difChain}
\end{figure}

A chain that differentiates \emph{correctly} (starting with a link at sepals) is the one whose first links specialize to sepals, next links specialize to petals, then stamens and carpels at the end (i.e. it recovers accurately the spatial configuration of the flower during cell fate determination). Recall that an undifferentiated chain is a radial longitudinal section of the meristem, so that the differentiated (or specialized) chain will be a radial longitudinal section of the flower (see figure (\ref{Fig:difChain})).

\subsection{Energy function}

We now define the energy of a specialization path and a function of energy of a chain that measures the energy it needs to differentiate.

\begin{defn}[Energy of a specialization path]
The energy of the specialization path from $c_1$, $SP(c_1)=\{c_1\to c_{2}\to\ldots \to c_{n}\}$ (where $c_{n}$ is the root of the tree we are on) is given by
$$E_{SP}(c_1)=\sum_{i=1}^n d(c_{i},c_{i+1})$$
\end{defn}

\begin{defn}[Energy of a chain]
The energy of a chain $C=[c_1,c_2,\ldots c_L]$ is given by
$$E(C)=\sum_{i=1}^L E_{SP}(c_i).$$
\end{defn}

Considering nature is efficient, we expect the chains that require the least amount of energy to specialize, to be the ones that differentiate correctly.
Hence, we will look for the chain of minimal energy. This is a non standard optimization problem that we will solve using a Genetic Algorithm.

\section{Genetic Algorithm }
\label{sec:3}

We wish to minimize the energy of the chains, that is, we wish to choose the chain that requires the least amount of energy to specialize. For this we design and implement a Genetic Algorithm. Note that a chain with more than $13$ links ($13$ being the number of components in each link) is redundant, so it will not be optimal (one with more than $13$ links can always be reduced to one with $13$). We can thus, limit our search to chains with $L=13$.

\subsection{Individuals}

An individual of the genetic algorithm together with an initial condition (a starting link) will determine a chain. The starting link of each individual will be a leaf on sepals tree $c_S$ (same value for every individual). From there we will generate the rest of the links by copying the previous one and modifying a randomly chosen component (if it was a $1$ it will become a $0$ and viceversa). We can represent each individual as

$$I_i=[p_1,p_2,\ldots,p_{12}]$$
where $p_i\in[1,13]$ represents the component that will be modified in the following link, so that the chain $C_i$ generated by $I_i$ will be

$$C_i=[c_S,c_1,\ldots,c_{13}]$$

where $c_1(j)=c_S(j), j\neq p_1$, $c_1(p_1)=mod(c_S(p1)+1,2)$ and in general $c_i(j)=c_{i-1}(j), j\neq p_i$ and $c_i(p_i)=mod(c_{i-1}(p_i)+1,2)$ (every component but one will be the same between neighbouring links, so that $d(c_i,c_{i+1})=1$.

We start by randomly generating a population of $N$ individuals. $N$ will be a fixed number throughout the execution of the algorithm.

\subsection{Evaluation and stopping criteria}

Having generated a whole population we will evaluate each chain $C_i$, that is, we will compute its energy $E(C_i)$ as defined in the previous section. Its fitness  will be a normalized value, inversely proportional to its energy value in such a way that the lower their energy is, the closer their fitness will be to $1$:
$$fit(I_i)=\frac{\alpha}{E(C_i)}$$
where $\alpha$ is a normalization function.

Once the whole population is evaluated we can proceed to apply the genetic algorithm operators (selection, crossover and mutation). These will generate a new population that will again be evaluated. The whole process will stop when we reach a minimum. Since we don't know the exact value of the minimum, the iterations will stop once the difference between the fitness values of one generation to the next is smaller that a given value $\epsilon$.

\subsubsection{Selection}

Once we have a set of individuals (a population) $P=[I_1,\ldots, I_N]$ that have been assigned a fitness value, we will select the individuals that will survive on to the next generation. We use the \textit{roulette wheel selection} method, a fitness proportionate selection, where each individual has a probability proportional to its fitness value to be chosen. Note that, for the next generation, each individual might be selected more than once, and there might be individuals that are not selected at all. To do this, we start by sorting the individuals by their descending fitness values. The accumulated normalized fitness values are computed (the accumulated fitness value of an individual is the sum of its own fitness value plus the fitness values of all the previous individuals). The accumulated fitness of the last individual should be $1$.
A random number $m\in[0,1]$ is chosen. The selected individual is the first one whose accumulated normalized value is greater than $m$. We repeat this process until $p$ individuals are chosen, so that the next generation is complete (the number of individuals in each generation will remain fixed through out all the iterations).

\subsubsection{Crossover}
\begin{figure}[h!]
\centering
  \includegraphics[width=16cm]{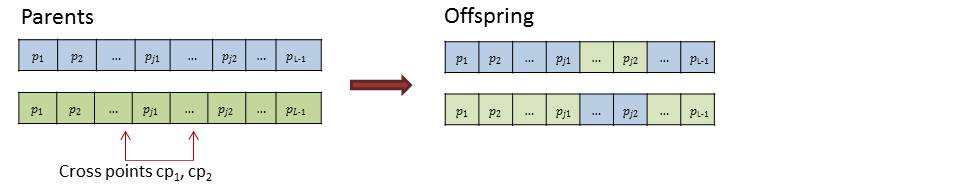}\\
  \caption{Crossover}\label{figAGCO}
\end{figure}

We use a two point crossover. For this, we select two individuals $I_i$ and $I_j$ using the selection operator. We will randomly choose two different crossover points $cp_1,cp_2 \in[1,L-1]$.
The two individuals will be combined forming two new ones: $I_i^n$ and $I_j^n$. The first part of $I_i^n$, up to the crossover point $cp_1$ and from $cp_2$ to the end, will be copied from $I_i$, the second part (from $cp_1$ until $cp_2$), will be copied from $I_j$. Conversely, the first part of $I_j^n$ (up to $cp_1$) and from $cp_2$ to the end will be copied form $I_j$ and the second part from $I_i$ (see figure (\ref{figAGCO})).
The two new individuals $I_i^n$ and $I_j^n$ each define new chains $C_i^n$ and $C_j^n$ that will substitute the old ones. We will have a fixed percentage of individuals that will be combined, in our work, we found experimentally that the best value was $80\%$.

\subsubsection{Mutation}
\begin{figure}[h!]
\centering
  \includegraphics[width=16cm]{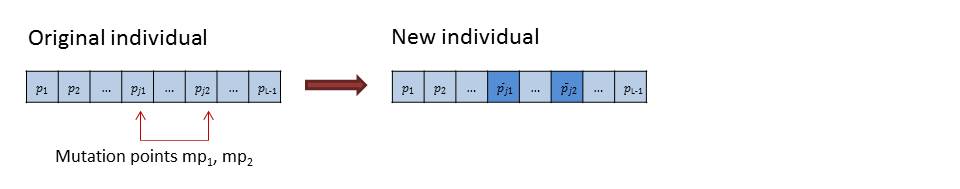}\\
  \caption{Crossover}\label{figAGM}
\end{figure}

The mutation operator is used to maintain genetic diversity from one generation to the next. It allows us to avoid local minimums when working with optimization problems. We will select one individual $I_i$ using the selection operator. We will randomly choose two mutation points $mp_1, mp_2 \in[1,L-1]$. We will replace the value of the states $mp_j$ in $I_i$ by a randomly chosen one between $1$ and $13$ (see figure (\ref{figAGM})). This will produce a new individual (with two mutated components):
$$I_i=[p_1,p_2,\ldots,p_{12}]$$
that defines a new chain that replaces the original one. As with crossover, we select a percentage of individuals that are mutated. We found experimentally $20\%$ to be the best choice.

\subsubsection{Elitism}
To make sure the we don't loose the best individual in each generation we apply elitism. We will find the individual with the highest fitness value, and copy it to the next generation with out modifying it. This will guarantee that the best individual in each generation is at least as good as the one in the previous one.

\subsubsection{Results}
We use the following parameters in our algorithm:
\begin{itemize}
  \item[1)] Length of the chain (number of undifferentiated cell types in a chain is $L=13$
  \item[2)] Maximum number of possible iterations allowed: $MaxG=200$
  \item[3)] Crossover probability: $cp=0.8$
  \item[4)] Mutation probability: $cp=0.2$
  \item[5)] Number of individuals per generation: $I=100$
  \item[6)] Elitism: $E=2$
\end{itemize}

After running the genetic algorithm an optimum solution is always found in less than $140$ iterations (generations). We show an execution where the minimum was found in $119$ iterations and the starting point is $0101100000000$ (a leaf in the sepals tree).

The chain with the lowest energy and the tree that each state belongs to, is shown in  table (\ref{tab2}).

\begin{table}[H]\label{tab2}
\caption{Minimum energy chain: first column shows the state of each cell type in the chain, the second one the tree it belongs to and the third one its specialization energy.}
\ \\
\centering
\label{tab:2}
\begin{tabular}{|c|c|c|}
   \hline
    State & Tree & Specialization Energy \\
    \hline
    0101100000000 & S & 12 \\
    0110110001100 & S & 1 \\
    0110110001110 & P & 0 \\
    1110110001110 & P & 1 \\
    0110110001110 & P & 0 \\
    0010110001110 & P & 1 \\
    0010110001111 & P & 17 \\
    0110110001111 & P & 0 \\
    0110110101111 & T & 2 \\
    0100110101111 & T & 1 \\
    1100110101111 & T & 0 \\
    1100110101101 & T & 1 \\
    1100110101100 & C & 0 \\
    \hline
    Total energy: & & E=36\\
    \hline
  \end{tabular}
\end{table}
\ \newline
where S, P, T, C correspond to Sepals, Petals, Stamens and Carpels respectively.

As we can see, this minimum energy chain traverses the trees in the correct orden, that is, following the flower's architecture which shows that we are correctly modelling it.

\section{Concluding remarks}
\label{sec:concl}

We are presenting a novel approach to analyze Genetic Regulatory Networks that allows us to unfold the richness and structure within them. In this case we used them to study the Genetic Regulatory Network of \textit{Arabidopsis thaliana} during cell fate determination and correctly recovered its architecture.

Genetic Regulatory Trees allow us to use tools from graph theory, finite field dynamical systems and finite field arithmetic to analyze the Genetic Regulatory Network behind them and to understand the principles that govern these networks better, which is our next goal. This method is general enough to be used in any GRN.

\newpage

\clearpage
\bibliographystyle{amsplain}
\bibliography{references}

\end{document}